# The Evolution of Zero Trust Architecture (ZTA) from Concept to Implementation

Md Nasiruzzaman
School of Computing & Data Science
Xiamen University Malaysia
Malaysia
PCS2411001@xmu.edu.my

Maaruf Ali
[0000-0001-9906-5004]
Faculty of Computer Science & IT
Universiteti Metropolitan Tirana
Albania
maaruf@ieee.org

The Doctoral College
University of Wales Trinity Saint David
Wales, UK

Iftekhar Salam
[0000-0003-1395-4623 ]
School of Computing & Data Science
Xiamen University Malaysia
Malaysia
iftekhar.salam@xmu.edu.my

Mahdi H. Miraz
[0000-0002-6795-7048]
School of Computing & Data Science
Xiamen University Malaysia
Malaysia
m.miraz@ieee.org

School of Computing,
Faculty of Arts,
Science & Technology
Wrexham University, UK

Faculty of Computing,
Engineering & Science
University of South Wales, UK

*Abstract*—Zero Trust Architecture (ZTA) is one of the paradigm changes in cybersecurity, from the traditional perimeter-based model to perimeterless. This article studies the core concepts of ZTA, its beginning, a few use cases and future trends. Emphasising the always-verify and least privilege access, some key tenets of ZTA have grown to be integration technologies like Identity Management, Multi-Factor Authentication (MFA) and real-time analytics. ZTA is expected to strengthen cloud environments, education, work environments (including from home) while controlling other risks like lateral movement and insider threats. Despite ZTA's benefits, it comes with challenges in the form of complexity, performance overhead and vulnerabilities in the control plane. These require phased implementation and continuous refinement to keep up with evolving organisational needs and threat landscapes. Emerging technologies, such as Artificial Intelligence (AI) and Machine Learning (ML) will further automate policy enforcement and threat detection in keeping up with dynamic cyber threats.

## I. INTRODUCTION: THE SHIFT TO ZERO TRUST (ZT)

Zero Trust Architecture (ZTA) has been a modern cybersecurity strategy, harnessed from the shortcomings of traditional perimeter-based security models. Traditional models depend on static network-based perimeters, which are proving quite inadequate in the present environment due to cloud services, remote workforces and increasingly blurred network boundaries [1]. These traditional approaches are based on the assumption that, once a user/device is inside the network, it can be trusted with broad access to resources [2]. This implicit trust is a big weakness since adversaries that are determined can breach network perimeters through different techniques such as phishing, supply chain attacks and exploitation of malicious insiders.

ZTA overcomes these limitations by taking the focus of security from network segments to users, assets and resources. Instead of trusting devices or users based on their physical or network location, ZTA requires that all access requests be always verified [2][3], as if they were coming from outside the network. The philosophy of "never trust, always verify" is deeply rooted in the approach [4] of ZTA.

This is a very fundamental principle that introduces a great shift from trust by implication to explicit verification. In the ZTA framework, no user, device or application is trusted inherently [5]. Each user or device must pass through the standard authentication and authorisation process before accessing data, applications or resources. This shall be based on verification related to various attributes, which include the identity of a user, the posture (i.e. security status and compliance with an organisation's security policies), of the device and the context of the request for access [6]. Through constant verification of the elements mentioned, ZTA ensures that only legitimate users and devices are granted access to sensitive resources.

The shift to ZTA is not an easy transition but a journey; it requires a complete rethink of the existing security practices. It calls for designing a more secure architecture that should be consolidated and flexible, without hindering operations or compromising security. This study reviews other researchers' and institutions' contribution to the development and implementation of ZTA over time and anticipates a possible trend for the future. This paper provides a starting point in order to present a high-level view of the journey of ZTA.

## II. Beginning of ZT: Ideas and Key Figures

The concept of ZTA really grew from ideas and contributions of those who challenged the traditional views of security. Amongst all, John Kindervag from Forrester Research is considered one of those seminal figures who formalised the core ideas of Zero Trust (ZT) in 2010 [2]. His work was called a direct response to inadequate perimeter-based security models and their inability to address up-to-date insider threats effectively.

Kindervag's original concept was radical: trust must be removed as a vulnerability [7]. He said the traditional approach of implicitly trusting users and devices inside a network was a critical flaw, leaving organisations vulnerable to both external attackers who breached the perimeter and malicious insiders. Kindervag's model was more a significant change to a model that suggested no default user, nor device and application should be trusted in regards to their location or network. This has found manifestation in the now-core principle in the ZTA framework. Kindervag also proposed three key principles for ZT security:

1. All sources must be verified and secured;
2. Access control must be limited and strictly controlled;
3. All network traffic must be inspected and logged.

His work emphasised the need to move away from a network-centric view of security to a more granular, resource-centric approach that focuses on data, users and devices.

The concept of de-perimeterisation, i.e. limiting the implicit trust based on network location, also gave birth to ZTA. For example, one such concept of de-perimeterisation was publicised by the Jericho Forum back in 2004 [8]. De-perimeterisation recognises risks from depending on single, static defences over a large network segment. This idea further matured and evolved into the bigger concept of ZT.

Another strong influence on the development of ZTA was Google's "BeyondCorp" project [9][10]. Starting in 2011, Google began building the BeyondCorp project. BeyondCorp was not just some theoretical exercises; rather, it was an actual implementation of ZT principles in a large, complex enterprise environment. It was a proof that a large organisation could work without a traditional, trusted network perimeter. BeyondCorp focused on access control and device accounting. It combined existing security technologies, such as access proxies and single sign-on, with a focus on application control rather than network control. The BeyondCorp project aimed to eliminate the privileged perimeter and migrate corporate applications to the internet for remote access. It would establish a system whereby access to resources was not based on network location but on the identity and security posture of users and devices. The project labelled devices and users with a dynamic trust level that it used in access control decisions. By 2017, it was fully deployed across the Google office network.

BeyondCorp became one of those proof-of-concepts to show how to move an organisation from a perimeter-based security model to one in which every user and device authenticates directly to the resource being accessed. Google's experience with BeyondCorp served as proof that a zero-trust model could actually be implemented successfully within an organisation as large and complex as Google. This helped serve as a practical guide for other organisations considering moving to ZTA. Most opportunistic, the success of BeyondCorp proved to be more than precise when remote work during the COVID-19 outbreak turned out to be the norm [10].

The birth origin of ZTA lies within the theoretical framework developed by John Kindervag - an explicit highlighting of the fragility of implicit trust. John Kindervag's ideas were put into practice through Google as the BeyondCorp project. These contributions have moulded ZTA into a holistic security strategy that overcomes the drawbacks of traditional security models with respect to modern, dispersed environments.

## III. Core Principles and Tenets of ZTA

ZTA is based on a set of core principles and tenets that work together to improve security by removing implicit trust and relying on explicit verification. While these vary slightly between sources, they all tend to come together on one common goal: protecting resources based on the assumption that no user, device or application can be trusted inherently. These tenets are not just abstract concepts but real design and implementation guidelines that would drive a robust security framework. The following presents some of these pioneering concepts:

- **Data Centric Security**: in ZTA, data protection is put at the very core of its philosophy. It acknowledges that data is an asset to be secured, no matter where it resides or from where it is accessed. A data-centric security architecture starts with the identification of sensitive data and critical applications ZT is to be implemented [9]. This includes classifying data with attributes such as personally identifiable information (PII), other sensitive information and many more to enable proper protection irrespective of the efficiency of other security measures. Data tagging on creation/import can also help in data categorisation.

- **Principle of Least Privilege Access**: this principle is central to ZT, ensuring that users and entities are only granted the minimum level of access necessary to perform their tasks [1][5]. It states that a user or device should be granted the least privilege required to operate and perform the work. The principle of least privilege reduces both the accessibility and visibility, which in turn minimises the lateral movement of an account in case it gets compromised [2], reducing the damage that an attacker can cause on a network. This simply means giving access only to data and resources required by each user or device.
- **Micro-segmentation**: ZTA segregates the network into smaller zones and makes them much more secure [2]. The approach creates isolated segments with granular access controls, making it challenging for threats to move within a network. In place of depending on wide network access, micro-segmentation isolates both workloads and applications as a way of reducing attack surfaces and controlling the traffic flow between segments. This gives an advantage over traditional perimeter security, as smaller segments reduce the ability of the attackers to move laterally within the network [2]. A macro and micro segmentation policy can be designed around segmenting and isolating specific workloads.
- **Continuous Authentication and Authorisation**: in ZTA, user and device identities are verified at each access request. That is, authentication and authorisation should not be single-point-in-time events but a set of continuous processes that allows access to only trusted users and devices [1]. This will be achieved by considering various attributes of the subject, such as identity, location, time and device security posture, to assess access requests beyond simple credential verification. Continuous authentication reduces the chances of unauthorised access and ensures that access is granted dynamically based on real-time conditions [1].
- **Policy-Based Access Controls**: under ZTA, access should be granted by predefined policy, considering user identity, device posture, time, location and data sensitivity [11]. It therefore, allows an organisation to grant access that is granular and fine-grained to satisfy a business requirement or regulatory compliance demand [5]. The policies can thus be based on various factors and sensitivity of resources including data, users, devices, threats and regulatory requirements. The policies need to be correct, consistent, minimal and complete.
- **Real-time Visibility and Analytics**: ZTA emphasises the continuous monitoring of network traffic and user behaviour for anomaly detection [12]. This real-time visibility is crucial in the early identification of potential security incidents and thus enables the rapid response of security teams to threats [8]. By collecting and analysing data about asset security, network traffic and access requests, ZTA allows organisations to enhance their security posture by identifying potential threats and vulnerabilities.
- **Automation and Orchestration**: ZTA leverages automation and orchestration for consistent and fast application of security policies. Such capabilities have become key to managing the complexity of ZTA at scale [7]. Automation dynamically reconfigures security policies, simplifies access control and supplies end-to-end zero-trust enforcement.
- **Assume Breach**: a cardinal principle of ZTA is to assume that the breach has already happened. The paradigm shifts from mere breach prevention to also containment and mitigation [5]. Security measures have been set in place, which aim to minimise the impact by limiting lateral movements and forbidding unauthorised access to sensitive information. The National Institute of Standards and Technology (NIST) Incident Response Cycle in Fig. 1, below, highlights a comprehensive approach to managing security breaches, including preparation, detection, containment and post-incident analysis [5]. This cycle ensures that even if a breach occurs, an organisation must be able to detect, respond and recover efficiently.

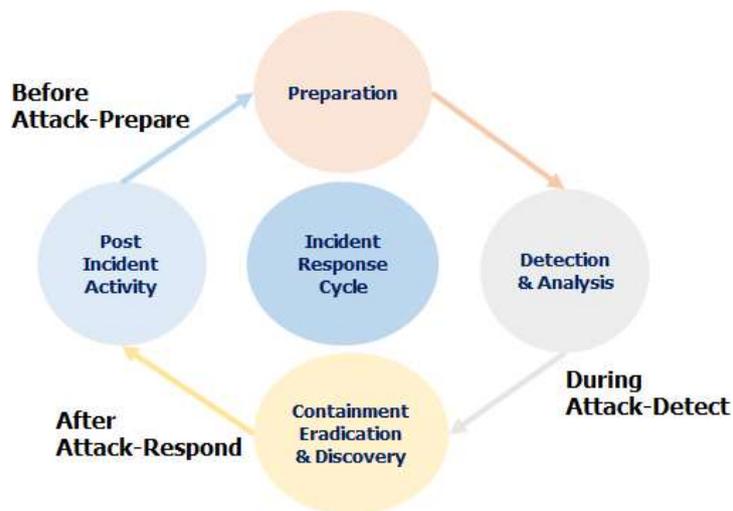

Figure 1. NIST Incident Response Cycle [5].

Notably, ZTA is not an architecture per se but rather a set of guiding principles for workflow and system design. The actual implementation can differ with the needs of an organisation, but core principles remain the same. Besides, it also requires current infrastructure and business process knowledge to map the access flows and develop the required security policies [13].

## IV. EVOLUTION OF ZTA IMPLEMENTATIONS

The traditional approach to security relies on a network perimeter, assuming that anything inside the network is trustworthy [2]. ZTA represents a fundamental shift in strategy for cybersecurity, from static, network-based perimeters to dynamic, granular focus on users, assets and resources. The reason behind this evolution is the increased complexity of the IT environment in modern times due to cloud technologies, remote workforce and Internet of Things (IoT) devices, amongst others, which have rendered traditional perimeter-based security models inadequate. According to Azad [14], the strength of social relationships amongst the users depends upon the services provided by the users and their IoT devices. With its fundamentally reimagined notion of how a network provides and handles trust and access, ZTA meets each challenge head-on. The conventional model of security still lives on the principle of "trust but verify": wide access to resources in the network is allowed as long as one believes that users and their devices are already trusted entities within it and the user authenticates through the network perimeter. In contrast, ZTA replaces this model in which no user and no device is trusted by default, regardless of where the location is or on which network it is connected [15]. Every access request shall be treated as potentially malicious and subjected to rigorous authentication, authorisation and continuous monitoring. It minimises lateral movement inside the network in case of a breach, hence mitigating the risk from both external and insider threats.

The most important point is that ZTA does not need to be deployed as some sort of forklift upgrade (major upgrade) of existing security infrastructure; rather, it deals with using existing security tools and technologies in an integrated and coordinated fashion. Traditional security tools such as Firewalls and Intrusion Prevention System (IPS)/Intrusion Detection System (IDS) add to the segmentation and detection of threats. Endpoint Detection and Response (EDR) systems will provide visibility and control to ensure the devices accessing are meeting the security posture requirements.

Security Information and Event Management systems gather security logs from different sources for analysis, anomaly detection, and incident response. Such tools, when integrated within a ZTA framework, elevate the overall security by providing the required data to make decisions and enforce actions at each access point.

Identity, Credential and Access Management (ICAM) and Multi-Factor Authentication (MFA) become central in ZTA [16]. ICAM enables the verification and management of user and device identities and their attributes. MFA adds an additional layer of security by requiring multiple forms of verification before access is granted, reducing the risk of unauthorised access due to compromised credentials. These measures ensure that only authorised entities have access to network resources. The attributes that are necessary for authorisation uniquely depend on the level of access granted, device hygiene and activities conducted within the environment. The ZTA also calls for resource access to be continuously monitored with dynamic risk-based assessments. All transactions and sessions shall be constantly reassessed in a ZTA: activities of users and devices are monitored continuously for suspicious behaviour. It allows real-time monitoring for changes in access control and enables the early detection of any potential threat. Decisions on access shall not be based on any static policy but will change dynamically with the continuous assessment of risk factors. This will be enabled so that the system learns the changing patterns of user behaviour, changes in device posture, and changes in the threat landscape-to grant access only under safe conditions and according to policy. It takes inputs from a combination of multiple factors at the user level, device hygiene and activity beyond mere authentication to determine dynamic confidence scores for assessing access.

Fundamentally, the shift in ZTA is toward dynamic security without static, perimeter-based approaches, but rather centred around resources with constant verification or assessment. This implies adherence to the principle of ZT, verification, utilisation of existing security tools, ICAM, MFA, continuous monitoring and risk-based assessment mechanisms that will further ensure protection for an organisation from most relevant cyber threats today.

## V. Implementation Challenges and Considerations

There are a number of significant challenges associated with implementing a ZTA and any transition to such an architecture needs to be carefully thought through. The transition from traditional, perimeter-based models to a ZTA is complex; therefore, it will be important to phase the implementation to avoid disrupting operations [1].

Among the main challenges is inherent complexity in transitioning from perimeter-based security. Whereas the traditional security models had trust on everything inside a network, ZTA enforces verification of every user, device and application to permit access. This represents a basic shift that needs a rethink of the current policies, workflows and technologies on security. It is going from broad network access to granular, context-aware, dynamic access controls. This transition demands a formidable upfront investment in resources and know-how.

The need to minimise business disruptions through phased implementation cannot be overlooked. Users are the most significant factor in the implementation prices because they can be prone to being compromised both inside and outside the network perimeters [16]. A complete, sudden implementation of a ZTA engenders major operational issues and thus the migration needs to take place in a piecemeal manner by implementing pilot programmes and further increasing the scope with gradually built confidence. A risk-driven, model-based migration process supports adapting existing infrastructures to this enhanced security model [4]. This type of phased approach allows an organisation to tune its policies, discover potential issues and effectively train users for a smoother transition. For instance, an organisation might begin with small-scale ZTA deployment for a small group of users or specific applications before expanding it to the entire enterprise as it requires a shift in mindset [17].

Another essential ingredient for successful implementation of ZTA is to have knowledge of all the assets, whether physical or virtual and all the subjects and their user privileges of the organisation's business processes. In the absence of effective understanding of the environment, defining appropriate access policies and implanting effective security controls would be quite challenging. The organisation has to conduct comprehensive asset surveys, data flow and workflow [16]. This generally includes "shadow IT" and unmanaged device identification, adding difficulty. The mapping of the flow of transactions also will be required for accessing certain data and their requirements toward security.

There are also several inherent vulnerabilities with a ZTA. Arguably the most critical of those concerns is a compromised control plane [18]: The policy engine and policy administrator sit inside the control plane. A compromised control plane allows attackers to access sensitive resources almost ubiquitously or disrupt critical functions en masse. Lessening this risk will come from extensive reviews of a control plane's attack surface, the redundancy of all components involved and regularly testing failover scenarios of each component.

There is performance overhead that could be associated with the implementation of ZTA. Continuous authentication and authorisation processes, together with micro-segmentation and encryption, may add latency and higher processing loads [18], impacting system performance. A study by Zanasi [2] marked that, a re-enrolment process caused a 100 ms spike in connection latency, while Windows clients experienced a 3-4 s delay. Another study by Pokhrel [15] showed an average ZTA operation delay about 2 s when compared to a baseline Federated Learning algorithm. The organisations must take careful consideration of the implications for performance, considering also the elastic properties of the ZTA in support of scaling. Such impacts should be very well understood upfront by using modelling, simulation, testing and pilot programmes before full deployment.

Implementing ZTA is not a quick process but rather a journey that requires incremental changes, process adjustments and technology solutions. It is not a "one size fits all" approach and must be modified to suit each organisation [13]. A ZTA implementation might require a few-year transition plan. It should include defining drivers and use cases, developing policies, designing the architecture, assessing technology readiness, conducting pilots, user training and phasing deployments. It is suitable for most enterprises to operate in a hybrid ZTA and perimeter-based mode. At the same time, investing in IT modernisation is crucial. A phased rollout may be considered to avoid disrupting business services and user experience [18]. It is important to migrate user groups gradually, starting with lower-risk groups. There is a lack of industry standards and concrete frameworks which can make implementation more challenging.

Organisations can transition to ZTA by using tools like Azure AD for identity management and conditional access. Similarly, Google's BeyondCorp model offers a ZT approach by authenticating all access requests. Whilst Cisco Zero Trust provides a framework for securing networks and endpoints [6]. These tools can help establish least-privilege access and continuous verification [13]. There is a perception that ZTA is only suitable for large organisations as it is perceived to require substantial investment. However, ZTA is just a set to guide principles that are suitable for organisations of any size.

## VI. ZTA IN VARIOUS ENVIRONMENTS AND USE CASES

ZTA is highly adaptable. Its application spans many environments and use cases, including but not limited to cloud computing, higher learning and remote work.

**Cloud Computing**: the demand for and utilisation of ZTA are very high in cloud computing [12]. Fig. 2, shows the key security risks associated with cloud storage which is a major component of cloud computing. These risks encompass a range of threats, from external attacks like DDoS (Distributed Denial-of-Service) and data breaches to internal vulnerabilities such as insufficient access management and insider threats.

Cloud environments are distributed, where the concept of traditional network perimeters does not exist; therefore, they are highly vulnerable to lateral movement. ZTA, thus, addresses these challenges in the following ways:
- Implementing strict Identity and Access Management (IAM) policies and procedures that ensure resources are accessed only by users, devices and applications that have a business need for such access [11].
- Network segmentation: this will limit the lateral movement of a malicious actor if there is a security breach [3].
- MFA to minimise unauthorised access.
- Continuous monitoring and anomaly detection: to spot security threats in real time and respond.
- Data-centric security: sensitive data identification and identification of critical applications.
- DLP (Data Loss Prevention) and DRM (Digital Rights Management) for Data Protection.
- Security of cloud-enabled IoT interactions since this interaction in nature is happening without any human intervention.
- This also enables organisation to protect their data as well as application even at remote server storing and processing.

**ZTA in Higher Education**: Higher Educational Institutions (HEIs) have their resources decentralised over a geographic location; hence it also gets great benefit from ZTA [17]. The environment of HEIs usually comprises the following features:
- Diverse user groups, including students, faculty and staff, with differing needs for access.
- A wide array of devices connecting to their networks, including personal devices.
- Applications and services are spread across multiple departments and sites.

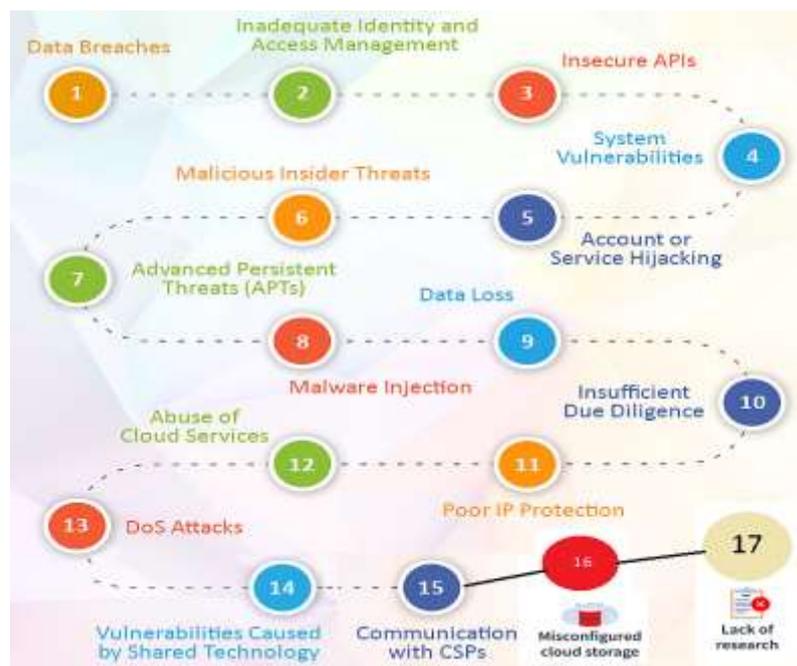

Figure 2.  Cloud Security Challenges[1] [12].

ZTA can contribute to the protection of the organisation through the implementation of granular access control, allowing users access to only what they should need and monitoring network traffic to detect and respond to potential threats. A ZTA framework can help enhance an organisation's security posture through the continuous monitoring, isolation, securing and control of all devices connected to the network [17].

**Remote Work Scenarios**: The principles of ZTA can be highly applicable to the remote work scenarios, which include BYOD (Bring Your Own Device). In remote work, there are a number of security challenges that are caused by such factors, such as the following:

   i. Users connect from untrusted networks, whereby they can be exposed to man-in-the-middle (MITM) attacks among other threats [19].
   ii. A wide variety of devices, some with inadequate security measures.
   iii. Data being accessed outside of the controlled environment of the enterprise network.

ZTA addresses these issues by:
- Assuming the local network connection is hostile and that all traffic is monitored [19].
- Requiring authentication and authorisation for all connection requests.
- Enforcing secure communications through encryption and source authentication.
- Implementing policies that limit access to resources based on the security posture of the device.

This is through the use of software-defined perimeters to enable secure remote access. ZTA also facilitates continuous authentication and dynamic access policies to grant access in regard to different factors like location, device posture and user behaviour. In this way, remote workers will be granted access to enterprise resources in a very secure manner with minimal security risks. It allows an organisation to use a cybersecurity posture that is more secure and dynamic.

## VII. THE ROLE OF AI IN ZTA

The future of ZTA will be very different with emerging technologies like ML and AI being integrated for advanced threat detection, response and policy management. AI and ML will be imperative in enhancing security through automation in ZTA [6][20].

- **AI-powered Anomaly Detection**: ML algorithms can look through volumes of data to develop a pattern of normal behaviour and then find deviations from it that might indicate a security threat [21]. Real-time threat detection is increasingly important given the sophisticated nature of evolving cyber-attacks [20]. This higher level of threat detection is crucial while implementing ZTA.
- **Automation of Policy Enforcement**: AI can also be used to automatically apply the ZTA policy by dynamically changing access controls in real-time, taking into consideration today's conditions and risk assessment [21]. It unburdens the workload from security personnel and makes the application consistent across the network.
- **SOAR (Security Orchestration, Automation and Response)**: AI is helpful in integrating various security tools and systems; hence control of different security systems is possible from one place [6]. This definitely improves threat response times and the control plane of ZTA, therefore making the security operations proactive and effective.
- **UEBA (User and Entity Behaviour Analytics)**: AI can track user and entity behaviour for anomalies [6]. It helps in continuous authentication processes and forms a significant part of ZTA.
- **Trust Scoring** can be helped through AI in the dynamic calculation of trust scores based on attributes like user behaviour, device hygiene and activities performed within the environment, which, in turn, will be used to make decisions regarding access in real time [16].
- **Automated Data Classification**: AI can classify and tag data automatically, which is a key element of ZTA [22].
- **Predictive Threat Analysis**: AI can predict attacks that are likely to occur, adapt the security measures proactively and help in strengthening ZTA's effectiveness.
- **Dynamic Policy Management** will also be a key focus. As environments and threats evolve, ZTA policies need to be agile and change. This would mean that the system updates the security policies and access controls automatically. AI can help in giving real-time insights into changes in the security environment that will, in turn, enable more dynamic and adaptive policies. This includes:
  - **Continuous Monitoring of ZTA Policy Effectiveness**, with feedback to refine those policies.
  - **Automation of Policy Adjustments** to real-time threat intelligence and changing user and device behaviour.

---

[1] veritis transcend, "Top 15 Cloud Security Threats, Risks, Concerns and their Solutions", 2025. Available: https://www.veritis.com/blog/top-15-cloud-security-threats-risks-concerns-solutions/ [accessed: 9th January, 2025].

Single view of the applied policy across the enterprise enables better coordination and understanding of how changes in one area will affect other areas. ZTA can integrate AI and ML to make it more robust, adaptive and efficient to face the dynamic cyber threat landscape.

## VIII. FUTURE TRENDS IN ZTA

*1) Increased Adoption Across Industries*

Organisations across various sectors such as in finance, the healthcare and governmental agencies are increasingly adopting ZT principles to enhance their security postures against the dynamic and evolving threats.

*2) Integration with Cloud Security*

As more businesses migrate to cloud environments worldwide, ZTA will be fully integrated with cloud security frameworks [23], enabling the seamless control and monitoring of data access across heterogenous (hybrid and multi-cloud) infrastructures.

*3) AI and ML Enhancements*

The use of AI [24] and ML will be critical in fully automating threat detection and near real-time response, allowing for more rapid dynamic adjustments to security policies based on immediate user behaviour and threat intelligence.

*4) Identity and Access Management (IAM) Evolution*

Enhanced IAM [25] solutions will be developed to support fine-grained access controls, utilising biometrics, MFA and continuous authentication methods.

*5) Micro-segmentation*

Organisations will increasingly employ micro-segmentation [2][26] techniques to limit lateral movement within networks thus reducing the attack surface and containing potential breaches.

*6) Zero Trust Network Access (ZTNA) Expansion*

ZTNA will expand as a critical component of remote work strategies, providing secure access to applications without exposing the entire network to external threats.

*7) Regulatory Compliance and Standards*

Compliance requirements will continue to drive the adoption of ZTA frameworks. The ZT interaction protocols are already part of the Level 1 Web3 technology stack [27], as shown in Fig. 3, below.

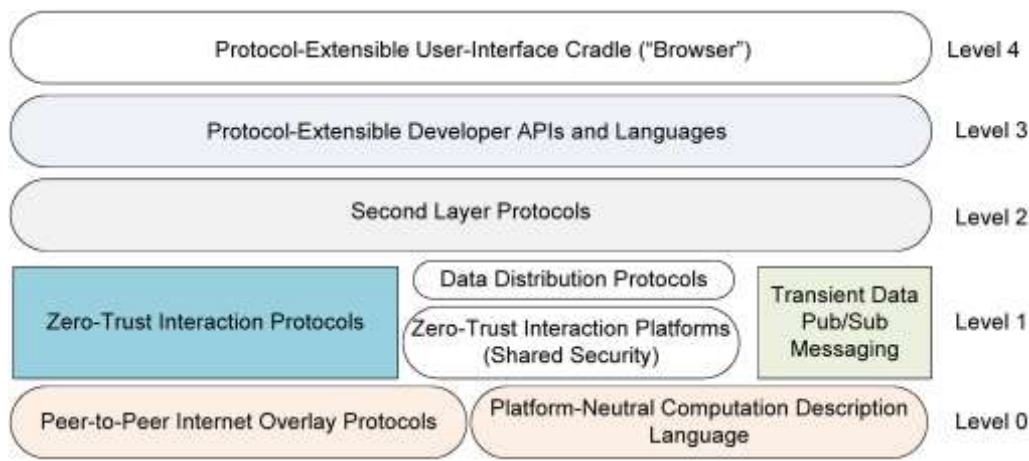

Figure 3. Web3 Technology Stack [27].

*8) Focus on Data-Centric Security*

Future implementations will prioritise protecting data itself, ensuring that sensitive information remains secure regardless of its location or the devices accessing it.

*9) Collaboration and Interoperability*

Increasing collaboration between security vendors to create interoperable solutions will be essential for implementing ZTA across diverse environments.

*10) User Education and Awareness*

As ZTA becomes ubiquitous, organisations will invest more in user education, awareness programmes - emphasising the importance of security hygiene and the principles of ZT.

## IX. CONCLUSION

ZTA represents a significant evolution in cybersecurity from the traditional perimeter-based models to more dynamic and granular approaches. ZTA is not an implementation but rather a continuous journey that requires steps in an incremental manner, complemented by ongoing adaptation. The key points of this evolution emphasise careful planning, a phased approach to implementation and continuous improvement.

Implementation of ZTA involves a bottom-up reassessment of all security policies and workflows. The first important thing to know in the implementation of ZTA is that it requires detailed knowledge of all assets, users and business processes so that granular access controls can be implemented to reduce the risk of lateral movement [8]. While ZTA offers the strongest form of security today, attention needs to be turned towards its weak points, one of which may include compromises in the control plane. There is continuous authentication and authorisation, but one has to take into account the performance overhead.

ZTA is not only versatile but also applicable in various cloud computing environments, higher learning institutions and even in some remote work environments. This is because its principles help in allowing secure access to resources across diverse and decentralised settings [28]. Future trends in the application of ZTA include increased integration with emerging technologies like AI and ML for threat detection, policy automation and dynamic policy management.

The most important criticism, however, is to keep in mind that ZTA is not mature yet. Final maturity of ZTA requires further research and more practical experience. Implementation of ZTA should be continuously monitored, assessed and refined against changes in the threat landscape and evolution of business needs. ZTA will continue to evolve as a critical strategy for organizations seeking to mitigate risks in an increasingly complex cybersecurity landscape. These trends will shape its future implementation and effectiveness.